\documentclass[a4paper,11pt]{article}
\pdfoutput=1 
\usepackage{jheppub, bm, color} 
\usepackage[T1]{fontenc}
\usepackage{amssymb,amsfonts,slashed,amsthm,amsmath,graphicx, soul}
\title{Inflation in multi-field random Gaussian landscapes}

\author{Ali Masoumi,}
\author{Alexander Vilenkin,}
\author{Masaki Yamada}

\affiliation{Institute of Cosmology, Department of Physics and Astronomy, 
Tufts University, Medford, MA  02155, USA}

\emailAdd{ali@cosmos.phy.tufts.edu}
\emailAdd{vilenkin@cosmos.phy.tufts.edu}
\emailAdd{Masaki.Yamada@tufts.edu}

\def\({\left(}
\def\){\right)}
\def\[{\left[}
\def\]{\right]}
\def\det{{\rm det}}

\def\pot{U}
\def\field{\phi}

\def\hess{\zeta}
\def\grad{\eta}
\def\thi{\rho}
\def\Tr{{\rm Tr}}

\def\lmk{\left(}
\def\rmk{\right)}
\def\lkk{\left[}
\def\rkk{\right]}
\def\dd{{\rm d}}

\newcommand{\eq}[1]{Eq.~(\ref{#1})}
\newcommand{\beq}{\begin{eqnarray}} 
\newcommand{\eeq}{\end{eqnarray}}
\newcommand{\bel}[1] {\begin{equation}\label{#1}}
\newcommand{\beal}[1] {\begin{eqnarray}\label{#1}}
\newcommand{\be}{\begin{equation}}
\newcommand{\ee}{\end{equation}}
\newcommand{\bea}{\begin{eqnarray}} 
\newcommand{\eea}{\end{eqnarray}}
\newcommand{\abs}[1]{\left\vert#1\right\vert}

\def\del{\partial}

\abstract{
 We investigate slow-roll inflation in a multi-field random Gaussian landscape.  The landscape is assumed to be small-field, with a correlation length much smaller than the Planck scale.  Inflation then typically occurs in small patches of the landscape, localized near inflection or saddle points.  We find that the inflationary track is typically close to a straight line in the field space, and the statistical properties of inflation are similar to those in a one-dimensional landscape.  This picture of multi-field inflation is rather different from that suggested by the Dyson Brownian motion model; we discuss the reasons for this difference.  We also discuss tunneling from inflating false vacua to the neighborhood of inflection and saddle points and show that the tunneling endpoints tend to concentrate along the flat direction in the landscape.

}

\begin{document}
\maketitle
\flushbottom

\section{Introduction} 
\label{sect:Intro}

String theory combined with inflationary cosmology has led to the picture of inflationary multiverse, populated by a multitude of vacua with diverse properties.  (For a review of multiverse cosmology and references to the literature see, e.g., \cite{Linde}.)
The vacua are represented by minima in the potential energy landscape, and transitions between different vacua occur by quantum tunneling through bubble nucleation.  All positive-energy vacua are sites of eternal inflation.  In addition, a realistic landscape should include regions allowing slow-roll inflation with $\gtrsim 50$ e-folds, leading to a low-energy vacuum like ours.  The expected number of vacua in the string landscape is enormous, so predictions in this kind of model should necessarily be statistical.  One may hope that the large number of vacuum states will make the statistical predictions sharp and simplicity will eventually emerge from the complex physics of the multiverse.

A natural first step is to study the statistics of a simple landscape described by a random Gaussian potential $\pot (\boldsymbol \field)$ in an $N$-dimensional field space $\phi_i ~(i=1, ..., N)$.  This approach has been adopted in much of the recent literature (see, e.g., \cite{Tegmark,Easther,Frazer,Battefeld,Yang,Bachlechner,Wang,MV,EastherGuthMasoumi,MVY1,MVY2}).  String theory suggests that the number of fields $N$ should be rather large, $N\gtrsim 100$, so one can use the properties of random Gaussian fields in the large-$N$ limit.  
It should be noted that a random Gaussian field does not reflect some qualitative features of the string landscape.  For example, the moduli potential in string theory should have decoupling limits, where the potential goes to zero. Kahler moduli may also have runaway instabilities~\cite{Dine:1985he, Bachlechner:2016mtp}. dS vacua are much more difficult to construct than AdS vacua in string compactifications -- which may or may not be represented by a random Gaussian potential with a constant term.  We also assume canonical kinetic terms for the moduli, which is generally not so in string theory.  A random Gaussian field should not therefore be regarded as a realistic model of the string landscape.  We believe, however, that understanding this model is an important first step, before the effects due to deviations from randomness or Gaussianity can be investigated.

In this paper we shall focus on so-called small-field landscapes, where the correlation length $\Lambda$ of the potential $\pot (\boldsymbol \field)$ is small compared to the reduced Planck scale, $\Lambda\ll M_{\rm Pl}$.  The conditions for slow-roll inflation in such a landscape are rather restrictive.  Inflation can typically occur in the vicinity of saddle points or inflection points of the potential \cite{LindeWestphal} (we shall specify the precise conditions in Sec. 2).  An estimate of the probability of inflation was attempted by Yang in Ref.~\cite{Yang}, with some {\it ad hoc} assumptions about the distribution of the field values after tunneling and about the attractor regions around inflection and saddle points that lead to inflation.  A different approach to the problem, using the random matrix theory, was initiated by Marsh {\it et al} in Ref.~\cite{Marsh} and further developed in \cite{Dias:2016slx,Freivogel,Westphal,Wang2,Marsh2}.  These authors noted that in order to deduce the inflationary properties of the landscape one only needs to know the potential in the vicinity of the inflationary paths. Furthermore, they conjectured that the evolution of the Hessian matrix $\zeta_{ij}=\partial^2 U/ \partial{\phi_i} \partial{\phi_j}$
along a given path in the landscape is described by a stochastic process that they specify (Dyson Brownian Motion, or DBM \cite{Dyson}).  This process is known to drive the Hessian distribution towards that of the Gaussian Orthogonal Ensemble (GOE).  With this assumption, the authors of \cite{Marsh,Dias:2016slx,Freivogel,Marsh2} have reached two 
major conclusions.  First, they found that inflation is generically multi-field, with a number of scalar fields participating in the slow roll.  And second, they found (in Ref.~\cite{Freivogel}) that inflation is far less likely than one might expect. Even if the slow-roll conditions are satisfied in a small patch of the landscape, the slope of the potential tends to rapidly steepen beyond that patch, cutting inflation short. Refs.~\cite{Marsh,Dias:2016slx,Freivogel,Marsh2} attribute these results to the fact that the statistics of Hessian eigenvalues in GOE is related to that of a gas of particles on a line interacting via a repulsive potential, resulting in `eigenvalue repulsion'.

This DBM method, however, has some problematic features.  The Hessian distribution in a random Gaussian landscape is significantly different from that in GOE; in particular, it gives a vastly larger density of minima \cite{Fyodorov,BrayDean,EastherGuthMasoumi}.  Some other problems with DBM have been pointed out in Refs.~\cite{Marsh,Freivogel,Wang2}.  The status of this method is therefore rather uncertain, and the conclusions it yields for inflation in the landscape should be taken with caution.

In two earlier papers \cite{MVY1,MVY2} we have developed precise analytic and numerical tools for studying inflation in a random landscape.  We applied these tools to the simplest case of a $1D$ landscape, where the potential depends on a single scalar field $\phi$.  In \cite{MVY1} we calculated the probability distributions for the maximal number of e-folds and for the spectral index of density
fluctuations, and in \cite{MVY2} we studied the distribution of scalar field values after tunneling and identified the attractor region around an inflection point that leads to inflation.  The purpose of the present paper is to extend some of these results to the case of a multi-dimensional landscape.

This paper is organized as follows. In the next Section we review some relevant properties of random Gaussian landscape models. In
Sec.~\ref{sec:multi} we study analytically the field dynamics during the curvature-dominated period after tunneling and during the subsequent slow-roll inflation. 
We identify an attractor region of initial condition after tunneling 
where slow-roll inflation can be realized.  In contrast to Refs.~\cite{Marsh,Dias:2016slx,Freivogel,Marsh2}, we find 
that the dynamics is effectively one-dimensional, without steepening, once the slow-roll conditions are satisfied.  We explain the difference of our results from the DBM approach in Sec.~\ref{sec:DBM}. 
In Sec.~\ref{sec:tunneling} we use an approximate analytic method to study instanton solutions and determine the initial conditions after tunneling.  We find that the initial values of the fields tend to concentrate along the flat direction in the landscape.  We verify this analytic treatment numerically in a simple model.  In most of the paper we focus on inflation near inflection points.  Analysis of saddle point inflation yields very similar results, as we briefly discuss in Sec.~\ref{sec:saddle}.  Finally, our conclusions are summarized and discussed in Section.~\ref{sec:conclusion}. 

\section{Random Gaussian landscape}\label{sec:landscape}

We consider slow-roll inflation in an isotropic $N$-dimensional random Gaussian landscape with a potential $U({\bm \phi})$.  The landscape is fully characterized by the average value ${\bar U}\equiv\langle \pot (\boldsymbol \field) \rangle$ and the correlation function 
\bel{Correlation}
\langle \pot (\boldsymbol \field_1) \pot(\boldsymbol \field_2)\rangle 
	- \bar{U}^2 
=  F (|\boldsymbol \field_1 - \boldsymbol \field_2|)=\frac1{(2\pi)^N}\int d^N {\bm k}\,P(k) e^{i{\bf k}\cdot (\boldsymbol\field_1-\boldsymbol\field_2)}~. 
\ee
Here, $k \equiv |\boldsymbol k|$ and angular brackets indicate ensemble averages.
Different moments of the spectral function $P(k)$ can be defined as  
\bel{sigmaDef}
	\sigma_{n}^2= \frac1{(2\pi)^N}\int d^N {\bm k} k^{2n} P(k)~. 
\ee

We assume that the potential $U({\boldsymbol \field})$ has a characteristic scale $U_0$ and a correlation length $\Lambda$ in the field space, with the correlation function $F (|\boldsymbol \field_1 - \boldsymbol \field_2|)$ rapidly decaying at $|\boldsymbol \field_1 - \boldsymbol \field_2| \gg \Lambda$.
We assume also that the ensemble average $\bar{U}$ is positive and is of the same order as $2\sqrt{N} U_0$, since otherwise most of the local minima of $U({\boldsymbol \field})$ would have a negative energy density.  However, we do not explicitly use this assumption in the paper. 

In this paper we focus on the case of a small-field landscape with $\Lambda\ll M_{\rm pl}$ and $U_0\ll M_{\rm pl}^4$, where $M_{\rm pl}$ is the reduced Planck mass, ($M_{\rm pl}\simeq 2.4 \times 10^{18} \ {\rm GeV}$).  Hereafter, we use the reduced Planck units ($M_{\rm pl} \equiv 1$) and assume $\Lambda \ll 1, ~ U_0\ll 1$. 

As a reference, we may consider a Gaussian-type correlation function defined as 
\be
F(\phi)=U_0^2 e^{-\phi^2/2\Lambda^2}. 
\label{correlation function}
\ee
In this case, the spectral function $P(k)$ is  
\be
\label{pk}
P(k)= U_0^2 (2\pi\Lambda^2)^{N/2} e^{-\Lambda^2 k^2/2}
\ee
and the moments are given by
\beq
\label{sigmaDef}
	\sigma_{n}^2 = \frac{2^n \Gamma \lmk n + \frac{N}{2} \rmk }{\Gamma \lmk \frac{N}{2} \rmk} 
	\frac{U_0^2}{\Lambda^{2n}}. 
\eeq
From this example, we expect $\sigma_n^2 \sim U_0^2 (N / \Lambda^2)^n$ in the large-$N$ limit for a generic correlation function. 
Although this estimate is valid in general, it should be noted that the Gaussian correlation function (\ref{correlation function}) is a very special case, in which the statistics of the potential minima is rather different from that for a generic correlator \cite{BrayDean}. In this paper, we do not use this correlation function but consider a generic case. We comment on the difference in Sec.~\ref{sec:DBM}.

\subsection{Inflation in a $1D$ landscape}

Here we review some results 
regarding one-dimensional random landscapes, which will be useful for our discussion later on.

The necessary conditions for slow-roll inflation in a one-dimensional inflaton potential $U(\phi)$ are 
\beq
\epsilon_s, \eta_s\ll 1, 
\label{1Dslowroll}
\eeq
where 
\beq
 &&\epsilon_s = \frac{1}{2} \lmk \frac{U'}{U} \rmk^2,
 \label{epsilon}
 \\
 &&\eta_s = \frac{U''}{U}. 
\label{eta}
\eeq
The typical values of the slow-roll parameters at a randomly chosen point in the landscape are $\epsilon_s \sim \eta_s \sim \Lambda^{-2}$.  In the small-field case $\Lambda\ll 1$, so typically $\epsilon_s , \eta_s \gg 1$ and inflation can occur only in rare regions where $U'$ and $U''$ are unusually small.  This is most likely to happen in the vicinity of an inflection point ($U'' = 0$) or of a local maximum of the potential ($U' = 0$).  On the other hand, the third derivative of $U$ in such regions needs not be particularly small and will typically be of the order $U''' \sim U/\Lambda^3$.

The range of the inflaton field where the slow roll conditions hold can be estimated from $|U'''| \Delta\phi \sim U$, or $\Delta\phi \sim \Lambda^3$.  This is much smaller than the correlation length $\Lambda$, and thus $U\approx {\rm const}$ within this range.  Furthermore, since $\Delta\phi\ll\Lambda$, the potential is well approximated by the first few terms in the Taylor expansion.  In the case of inflection-point inflation, we can write%
\footnote{
For consistency of notation with Refs.~\cite{MVY1, MVY2}, we use the notation $\eta$ for $U'(0)$. Note that it should not be confused with the slow-roll parameter $\eta_s$ in \eq{eta}. 

}
\beq
 U( \phi) = U + \eta \phi + \frac{1}{6} \rho \phi^3 
\eeq
with $\eta\rho >0$.  Here, $\eta = U'(0)$, $\rho = U'''(0)$ and the inflection point is at $\phi=0$.  Without loss of generality we can set $\eta,\rho <0$.\footnote{For $\eta\rho >0$ the potential has a local maximum and a minimum at $\phi = \pm (2\eta/\rho)^{1/2}$.  Saddle-point inflation is then possible at the local maximum.  We discuss this case in Sec.~\ref{sec:saddle}.}

The magnitude of density perturbations $\Delta_R^2$ and the spectral index $n_s$ are given by \cite{Baumann}
\beq
 &&\Delta_R^2 = \frac{1}{12 \pi} \frac{U^3}{\eta^2} = \frac{N_{\rm max}^4}{48 \pi^6} \frac{\rho^2}{U}, 
 \\
 &&n_s \simeq 1 - \frac{4 \pi}{N_{\rm max}} \cot \lmk \frac{\pi N_e^{\rm (CMB)}}{N_{\rm max}} \rmk, 
\eeq
where $N_e^{\rm (CMB)}$ ($\simeq 50-60$) 
is the e-folding number at which the CMB scale leaves the horizon. 
We also defined the maximal e-folding number as 
\beq
 N_{\rm max} \approx - \int_{- \infty}^\infty \dd \phi \frac{U(\phi)}{U'(\phi)} \approx 
 \pi \sqrt{2} \frac{U}{\sqrt{\eta \rho}}. 
\label{Nmax}
\eeq
The observed value of the spectral index ($n_s \simeq 0.97$) is obtained 
when $N_{\rm max} \approx 120$. 
The magnitude of density perturbation can be consistent with the observed value ($\Delta_R^2 \sim 4 \times 10^{-9}$) 
if we choose $U\sim 10^{-12} \Lambda^6$.

Apart from the conditions (\ref{1Dslowroll}), slow roll inflation requires appropriate initial conditions for the field $\phi$.  These conditions are determined by the instanton solution describing the bubble nucleation.  Inflation can occur only if the initial value of $\phi$ after the tunneling is sufficiently close to the inflection point.  It was shown in \cite{MVY2} that the corresponding attractor range of $\phi$ is
\beq
-\kappa (2U/3|\rho|) < \phi \lesssim U/|\rho|, 
\label{1Dattractor}
\eeq
where $\kappa \approx 24.8$ and $U/|\rho| \sim \Lambda^3$.  Furthermore, it was also shown in \cite{MVY2} that instanton solutions describing tunneling to a vicinity of an inflection point do exist if the potential at that point is sufficiently flat.  However, the ensemble distribution of the initial values of $\phi$ is rather broad, with a width $\sim (0.1 - 0.4)\Lambda$.  The distribution is more or less flat in this range, so tunnelings to the small attractor region have probability $\sim \kappa\Lambda^2$.  We will show in Sec.~5 that such tunnelings require a thin-wall bubble, with the potential $U$ at the inflection point nearly degenerate with that at the false vacuum.

\subsection{Taylor expansion around an inflection point}

In a multi-dimensional landscape, slow-roll inflation still requires a sufficiently flat region of the potential.  The corresponding conditions can be written as (e.g., \cite{Yang}) 
\beq
 &&\epsilon_s = \frac{\del_i U \del_i U}{2 U^2} \ll 1, 
 \\
 &&\eta_s = \sqrt{ \frac{\del_i U ( \del_i \del_j U ) ( \del_j \del_k U ) \del_k U }{\abs{\del_i U}^2 U^2} } \ll 1, 
\eeq
where we use Einstein's summation convention.%
\footnote{
These conditions are sufficient for slow-roll inflation. 
We disregard the special cases where inflation can occurs with weaker conditions.  
}
As in the $1D$ case, one can expect that inflation occurs in a small patch $|\Delta\boldsymbol \field | \ll \Lambda$; then the potential is well approximated 
by a cubic expansion
\bel{potential} 
U({\bm \phi}) = U + \grad_i \phi_i + \frac{1}{2} \hess_{ij} \phi_i \phi_j + \frac{1}{6}\thi_{ijk} \phi_i \phi_j \phi_k~, 
\ee
where $i,j,k = 1,2, ... , N$.  These expectations will be justified {\it a posteriori}.  The expansion coefficients in Eq.~(\ref{potential}) are $\grad_i \equiv \del U / \del \phi_i$, 
$\hess_{ij} \equiv \del^2 U / \del \phi_i \del \phi_j$, 
and $\thi_{ijk} \equiv \del^3 U / \del \phi_i \del \phi_j \del \phi_k$, with all derivatives taken at $\phi_i =0$. 
Note that the indices of $\hess_{ij}$ and $\thi_{ijk}$ have a symmetry under the interchanges of 
$i \leftrightarrow j \leftrightarrow k$. 
For example, the coefficient of $\phi_1 \phi_2^2$ is $(\rho_{122} + \rho_{212} + \rho_{221}) / 6 = \rho_{122}/2$.   The typical values of the expansion coefficients in (\ref{potential}) are $\grad_i\sim U/\Lambda$, $\hess_{ij}\sim U/\Lambda^2$, $\thi_{ijk}\sim U/\Lambda^3$.  Their probability distribution has been found in Refs.~\cite{Fyodorov,BrayDean,MVY1}.  

A multi-field hilltop inflation occurs near a stationary point where $\nabla U=0$.  The Hessian at this point must have one or several small negative eigenvalues, $|m_i^2| \ll U$, with other eigenvalues being positive and typically having their generic values.

A multi-field analogue of inflection-point inflation occurs near a point where the gradient of the potential is small and one of the Hessian eigenvalues is zero.  The latter condition can be stated as $\det \hess = 0$.  
We can choose the basis in the
$\phi$-space so that the matrix $\hess_{ij}$ is diagonal, 
\beq 
\hess_{ij} = m_i^2 \delta_{ij} , 
\eeq 
and the zero eigenvalue corresponds to $i = 1$:
\beq
m_1 =0. 
\eeq 
The other eigenvalues $m_a^2$ will typically have generic values, taken from a distribution that we shall discuss in the next subsection.   A typical eigenvalue is of the order $m_a^2 \sim \sqrt{N} U_0 / \Lambda^2$, while the smallest nonzero eigenvalues are $m_a^2 \sim U_0 / (\sqrt{N} \Lambda^2)$.   If one of these eigenvalues is negative, it would trigger a tachyonic instability and a long period of inflation would be impossible.  Hence we assume that the Hessian $\hess_{ij}$ has all but one positive eigenvalues. 
Here and hereafter, we use the notation that 
the subscript $a$ runs from $2$ to $N$ while the subscripts $i,j,k$ run from $1$ to $N$. 

The condition $\det \hess = 0$ specifies a codimension-1 surface in the field space.  When $|\nabla U|$ is small, we can find a nearby point on this surface where the gradient $\nabla U$ is directed along the 1-axis: 
\beq
\grad_i = \grad \delta_{i1}. 
\eeq
We shall refer to this point as the inflection point.  The potential near this point has the form of a groove running in the $\phi_1$-direction between the hills that surround it in the orthogonal $\phi_a$-directions.

In most of this paper we are going to focus on inflection-point inflation.  Saddle-point inflation can be analyzed in a very similar way; we shall discuss it briefly in Section 6.

\subsection{Hessian eigenvalue distribution}

Of particular interest is the distribution for the eigenvalues $\lambda_i =m_i^2$ of the Hessian matrix $\zeta_{ij}$.  This is given by the `semicircle law', 
\beq
\rho(\lambda)=\frac{2}{\pi b^2 N}\left( b^2 N -(\lambda-{\bar\lambda})^2\right)^{1/2} .
\label{Wigner}
\eeq
Here, $\rho(\lambda) d\lambda$ is the number of eigenvalues in the interval $d\lambda$, ${\bar\lambda} =N^{-1}\sum_i \lambda_i$ is the average eigenvalue, 
\beq
b^2=\frac{4\sigma_2^2}{N(N+2)},
\label{b}
\eeq
and $\sigma_2^2$ is the second moment of the correlation function, as defined in (\ref{sigmaDef}).
Eq.~(\ref{Wigner}) applies in the range $\abs{{\lambda} -{\bar\lambda}}\leq b \sqrt{N}$, with $\rho(\lambda) =0$ outside this range.

The quantity $b^2\sim U/\Lambda^2$ is the characteristic dispersion of the matrix elements $\zeta_{ij}$; it is independent of $N$ in the large-$N$ limit.  We note, however, that the width of the distribution (\ref{Wigner}) is greater than $b$ by a large factor $\sqrt{N}$.  This is due to the `eigenvalue repulsion' phenomenon.

Eq.~(\ref{Wigner}) with ${\bar\lambda}=0$ was derived by Wigner \cite{Wigner} as the eigenvalue distribution for a large random matrix.  In the case of a random Gaussian field, Bray and Dean \cite{BrayDean} showed that the Hessian eigenvalue distribution at stationary points is given by Eq.~(\ref{Wigner}) with the average eigenvalue ${\bar\lambda}$ related to the value of the potential, 
\beq
{\bar\lambda}=-\frac{\sigma_1^2}{N\sigma_0^2} \left(U-{\bar U}\right).
\label{barlambda} 
\eeq
For $U<{\bar U}$ the distribution is shifted towards positive values, and the entire distribution shifts to the positive domain when $U$ gets below certain critical value (defined by the condition ${\bar\lambda}=b\sqrt{N}$).  In this range of $U$, 
almost all of the stationary points of the potential are local minima.  Similarly, there is a positive critical value of $U$, above which {\b almost all of the} stationary points are local maxima.

The semicircle law (\ref{Wigner}) can also be used to describe the conditional eigenvalue distribution, under the requirement that all eigenvalues are greater than some $\lambda_*$ \cite{BrayDean}.  In this case, ${\bar\lambda}=\lambda_* + b\sqrt{N}$.  In particular, at an inflection point, where one eigenvalue is zero and the rest are positive, the distribution is given by (\ref{Wigner}) with 
${\bar\lambda}=b\sqrt{N}$.  

The semicircle law applies in the limit of $N\to\infty$, but for a finite $N$ it becomes inaccurate in small regions near the edges of the distribution.  Such edge corrections are important for the estimate of the smallest nonzero Hessian eigenvalue $\lambda_{\rm min}$ at an inflection point.  It can be shown that 
\beq
\lambda_{\rm min}\sim N^{-1/2} b
\label{lambdamin}
\eeq
and that the number of such eigenvalues is $\sim N^{1/4}$.  (Details of this analysis will be published elsewhere \cite{Masaki}.)  Thus, for $N\sim 100$ we can expect to have a few eigenvalues of magnitude $0.1~U_0/\Lambda^2$.

\section{Multi-field inflection-point inflation}\label{sec:multi}

After tunneling, the bubble has the geometry of an open FRW universe,
\beq
ds^2 = dt^2 -a^2(t)\left(d\chi^2+\sinh^2\chi d\Omega^2\right),
\eeq
with spatially homogeneous fields $\phi_i(t)$.
The evolution of $a$ and $\phi_i$ is described by the equations
\beq
\frac{{\dot a}^2}{a^2}=\frac{1}{3}\left(U({\bm \phi})+\sum_i \frac{{\dot\phi}_i^2}{2}\right)+\frac{1}{a^2},
\label{Friedmann}
\eeq
\beq
{\ddot\phi}_i +3\frac{\dot a}{a}{\dot\phi}_i + \frac{\partial U({\bm \phi})}{\partial \phi_i} =0,
\label{phiU}
\eeq
where dots represent derivatives with respect to $t$. 
The initial conditions at $t=0$ are given by 
\beq
a(0)=0,~~~{\dot a}(0)=1,~~~ \phi_i(0)=\phi_{i,0},~~~ {\dot\phi}_i(0)=0,
\eeq
where $\phi_{i,0}$ is determined from the instanton solution that describes the tunneling.

During a small-field inflation the potential (\ref{potential}) is nearly constant, $U({\bm \phi}) \simeq U ={\rm const}$, and
the Friedmann equation (\ref{Friedmann}) can be approximated as
 \beq
{\dot a}^2 = 1 + H^2 a^2,
\eeq
where $H^2 = U /3$.  The solution is the de Sitter space,
\beq
a(t) = H^{-1} \sinh (H t), 
\label{at}
\eeq
which gives $\dot{a}/a = H \coth (H t)$.  This shows that inflation starts at $t \sim H^{-1}$, after a brief curvature-dominated period.

\subsection{Starting with $\phi_a \approx 0$}\label{sec:phia=0}

Let us first consider the case when the initial values ${\bm \phi}_0$ are such that $\partial U / \partial \phi_a ({\bm \phi}_0)= 0$, while $\partial U / \partial \phi_1 ({\bm \phi}_0)$ is nonzero.  Then the field ${\bm \phi}$ starts rolling in the $\phi_1$-direction with $\phi_a \approx 0$, and we can expect inflation to be essentially one-dimensional, at least initially.  

\begin{figure}[t] %  figure placement: here, top, bottom, or page
   \centering
   \includegraphics[width=2.5in]{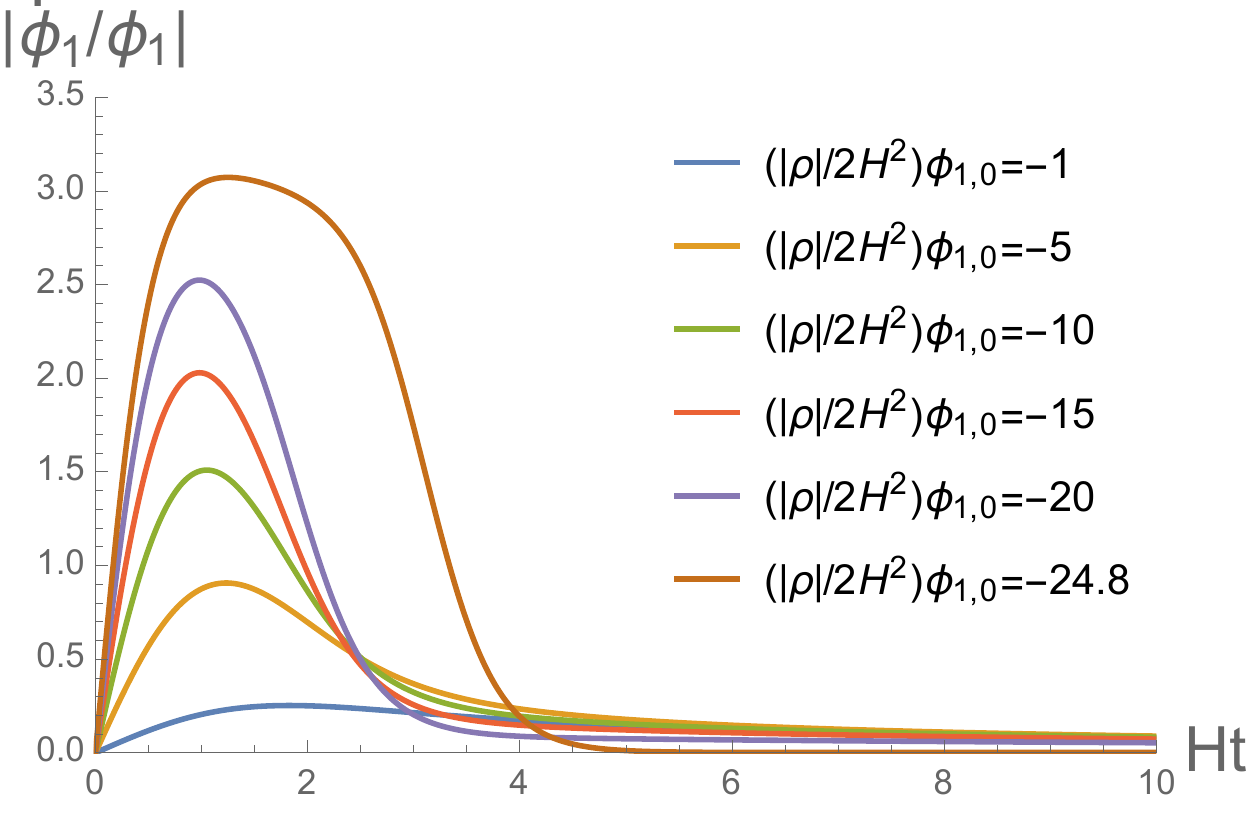} 
   \caption{
Plot of $\abs{\dot{\phi_1}/ \phi_1}$ as a function of time in units of $H^{-1}$. 
We assume $\eta = 0$. 
   }
   \label{fig:adiabatic cond}
\end{figure}

Neglecting $\phi_a$ and using the scale factor (\ref{at}) in Eq.~(\ref{phiU}), we obtain the following equation for $\phi_1(t)$
\beq
{\ddot\phi}_1 + 3{H} \coth ({H} t) {\dot\phi}_1 + \thi \phi_1^2 /2 + \grad  =  0, 
\label{phieq1}
\eeq
where we have introduced the notation $\thi_{111}\equiv \thi$. 
Hereafter we assume $\grad, \rho < 0$ without loss of generality.   As we mentioned in Sec. 2.1, the analysis of one-dimensional inflection-point inflation in Ref.~\cite{MVY2} has shown that 
$\phi_1$ does not overshoot the slow-roll region if its initial value is in the range (\ref{1Dattractor}),
\beq
-2\kappa H^2 / \abs{\thi} <\phi_{1,0} \lesssim H^2/ \abs{\thi}, 
\label{phi1range}
\eeq
where $\kappa \approx 24.8$. 
It was also shown in \cite{MVY2} that
in most of this range the last term in Eq.~(\ref{phieq1}) has negligible effect on the dynamics.  For later convenience, we numerically calculated $\abs{\dot{\phi_1}/ \phi_1}$ 
as a function of time for $\eta=0$.  The resulting plot in Fig.~\ref{fig:adiabatic cond} shows that this quantity does not exceed $3 H$. 

The fields $\phi_a$ with $a=2, ..., N$ are affected by the dynamics of $\phi_1$ 
because of the interaction terms.  The most important contribution comes from the term $\thi_{11a} \phi_1^2 (t) \phi_a$ in the potential, which introduces a force term in the field equation for $\phi_a$: 
\beq
 \ddot{\phi}_a + 3 H \coth (H t) \dot{\phi}_a + m_a^2 \phi_a 
 + \frac{1}{2} \thi_{11a} \phi_1^2 (t) = 0. 
\eeq
The effect of this term is to shift the minimum of the potential in the orthogonal directions to
\beq
 {\bar\phi}_a  (t) = - \frac{\thi_{11a} \phi_1^2 (t)}{2 m_a^2}. 
\eeq
The typical rate of variation of ${\bar\phi}_a (t)$ is $\sim\abs{\dot{\bar\phi}_a / {\bar\phi_a} }
= 2 \abs{\dot{\phi_1}/ \phi_1}$. 
From Fig.~\ref{fig:adiabatic cond}, we find that this rate is $\lesssim 6H$.   
On the other hand,
the oscillation rate of $\phi_a$ is $m_a \gtrsim N^{-1/4} \sqrt{U_0}/\Lambda$, where we have used the estimate (\ref{lambdamin}) for the smallest nonzero eigenvalue of the Hessian.
For $\Lambda \ll \sqrt{U_0 / H^2}/(6N^{1/4})$ ($\gtrsim 0.1$), we have $m_a\gg 6H$, 
where we have used $H^2 \simeq U / 3 \lesssim U_0/3$. 

This means that
the minimum of $\phi_a$ changes adiabatically, 
and thus oscillations of $\phi_a$ are not excited by the effect of interaction. 
We can then approximate $\phi_a \simeq {\bar\phi}_a (t)$ 
and obtain 
\beq
 \abs{ \frac{\phi_a}{\phi_1}} \simeq 
 \abs{\frac{\thi_{11a} \phi_1 (t)}{2 m_a^2} } \lesssim {\kappa}N^{1/2} \Lambda^2, 
\eeq
where we have used $\abs{\phi_1 (t)} \lesssim {\kappa} H^2 / \abs{\thi}$.  For $\Lambda\ll 0.1$ this gives $ \abs{{\phi_a}/{\phi_1}}\ll 1$, so the inflaton trajectory is approximately a straight line in the field space.

For moderately large values of $\Lambda \sim 0.1$, some low-mass modes $\phi_a$ may be excited
and the field trajectory may be significantly curved.  However, we shall see in the next subsection that oscillations of such modes are rapidly damped and the filed trajectory becomes straight during the slow roll, unless $\Lambda\gtrsim N^{-1/4} \sim 0.3$.

\subsection{Generic initial conditions}
\label{sec:generic}

Now let us relax the assumption that the gradient of the potential is aligned with the Hessian eigenvector with zero eigenvalue. In this case the gradient has nonzero components in directions orthogonal to $\phi_1$.  Let us call these components $\grad_a$.  For sufficiently small $\grad_a$, if we move
the origin of $\phi_{a}$ to $-\grad_a /m_a^2$, the gradient would be aligned with the $\phi_1$-
direction. Therefore, having a misalignment between the gradient and Hessian eigenvector is equivalent to choosing the fields $\phi_a$ with some displacement from their value
which minimizes the potential. Hence, we will study the field evolution with $\phi_{a,0}\neq 0$. Now $\phi_a$ will oscillate, and this may cause $\phi_1$ to move
fast and ruin the slow-roll conditions.  Our goal is to estimate the range of initial conditions which lead to a slow-roll inflation.  We shall first study the dynamics of $\phi_1$ and $\phi_a$ analytically under some plausible assumptions and then verify the results in a numerical example.

Suppose $\abs{\phi_{1,0}} \ll \abs{\phi_{a,0}}\ll\Lambda$.  The dynamics of $\phi_a$ are then mostly driven by their mass terms,
\beq
 \ddot{\phi}_a + 3 H \coth \lmk H t \rmk  \dot{\phi}_a + m_a^2 \phi_a = 0. 
\label{phiaeq}
\eeq
Focusing first on the curvature-dominated period, $t\ll H^{-1}$,
we can approximate $H \coth \lmk H t \rmk$ as $1/ t$. 
Then the solution of Eq.~(\ref{phiaeq}) is
\beq
 \phi_a (t) = 2 \phi_{a,0} \frac{J_1 (m_a t)}{ m_a t}, 
\label{phia}
\eeq
where $J_1 (z)$ is the Bessel function of the first kind. 

The field equation for $\phi_1(t)$ can be written as
\beq
{\ddot\phi}_1 +3\frac{\dot a}{a} {\dot\phi}_1 + \eta + \frac{1}{2} \rho_{1aa}\phi_a^2 =0,
\label{phi1eq}
\eeq
where we neglected $\phi_1^2$ compared to $\phi_a^2$. 
We also neglected the term proportional to $\rho_{11a}$, 
because $\phi_a (t)$ oscillates with a period much shorter than the 
typical time scale of $\phi_1$, so this term averages out to zero. 
With $y(t) \equiv {\dot\phi}_1(t)$, Eq.~(\ref{phi1eq}) takes the form
\beq
\frac{d}{dt}(ya^3) = 
-\eta a^3 
- \frac{1}{2} \rho_{1aa}\phi_a^2 a^3,
\eeq
and the solution is
\beq
y(t) = 
-\eta a^{-3}(t)\int_0^t dt' a^3(t')
 - \frac{1}{2} \rho_{1aa} a^{-3}(t)\int_0^t dt'a^3(t')\phi_a^2(t').
\label{ygeneral}
\eeq
The first term in (\ref{ygeneral}) is $\approx -\eta t/4$ for $t\ll H^{-1}$ and $-\eta/3H$ for $t\gg H^{-1}$.
We first disregard this term and take it into account later.

During the curvature dominated period, we have $a(t)\approx t$ and
\beq
y(t) = -\frac{2\rho_{1aa} \phi_{a,0}^2}{m_a^2 t^3} \int_0^t dt' t' J_1^2(m_a t')
      = -\frac{\rho_{1aa} \phi_{a,0}^2}{m_a^2 t} \left[ J_1^2(m_a t) -J_0(m_a t)J_2(m_a t)\right], 
\label{y2}
\eeq
where in the last step we used Eq.~(5.54(2)) in Ref.~\cite{GR}.  $\phi_1(t)$ can now be found from
\beq
\phi_1(t) = \phi_{1,0} + \int_0^t dt' y(t').
\eeq
This integral is calculated in Appendix~\ref{Bessel}, with the result   
\beq
\phi_1(t)=-\frac{\rho_{1aa}\phi_{a,0}^2}{2m_a^2} \left[1-J_0^2(m_a t) -2J_1^2(m_a t) +J_0(m_a t)J_2(m_a t) \right].
\label{Bessel result}
\eeq

Using the asymptotic forms of Bessel functions at small and large values of the argument, we find 
\beq
\phi_1(t) \approx \phi_{1,0} -\frac{1}{16} \rho_{1aa}\phi_{a,0}^2 t^2
\eeq
at $t \ll m_a^{-1}$ and
\beq
\phi_1(t)\approx \phi_{1,0} -\frac{\rho_{1aa}\phi_{a,0}^2}{2m_a^2} \left(1-\frac{4}{\pi m_a t}\right)
\label{phi1}
\eeq
at $t\gg m_a^{-1}$.

The effect of the linear term in the potential for $\phi_1$ (i.e., of the first term in \eq{ygeneral}) 
can be trivially taken into account by replacing 
\beq
 \phi_1(t) \to \phi_1(t) - \frac{\eta t^2}{8}. 
\eeq

Eq.~(\ref{phi1}) shows that the effect of the oscillating fields $\phi_a$ is to shift $\phi_1$ by the amount
\beq
{\cal S} = - \sum_a \frac{\rho_{1aa}\phi_{a,0}^2}{2m_a^2}, 
\label{shift}
\eeq
where we explicitly wrote the summation over $a$ ($= 2,3, \dots, N$). 
It also follows from (\ref{phi1}) that interactions with $\phi_a$ become unimportant at $t\gg m_a^{-1}$.  

During the curvature dominated period, the oscillation amplitude of $\phi_a$ decreases as $t^{-3/2}$.  This period may be followed by a period of slow-roll inflation, when the Hubble parameter is $H\approx (U/3)^{1/2} = {\rm const}$.  The field equation for $\phi_a$ is then
\beq
 \ddot{\phi}_a + 3 H \dot{\phi}_a + m_a^2 \phi_a = 0, 
\eeq
and its solution is
\be
 \phi_a(t) = C e^{-3H t/2} {\rm cos} \lmk m_a \sqrt{1- \zeta_a^2} t +\psi \rmk ~,
 \ee
where $\zeta_a \equiv {3 H}/{2 m_a} \lesssim N^{1/4} \Lambda$.  The constant amplitude $C$ and phase $\psi$ can be found by matching to the curvature-dominated regime.  For $\zeta_a < 1$ the fields $\phi_a$ rapidly decrease, becoming increasingly unimportant. 
The problem then reduces to that of a $1D$ landscape, discussed in Ref.~\cite{MVY2} and reviewed in Sec.~2.1.   We expect the condition $\zeta_a < 1$ to be satisfied, unless $\Lambda\gtrsim N^{-1/4} \sim 0.3$.

Since the inflationary dynamics is essentially one-dimensional, one can expect that the probability distribution for the maximal number of e-folds $N_{\rm max}$ is the same as in a one-dimensional landscape.  A detailed calculation in Appendix~\ref{sec:N_e} shows that this is indeed the case, 
and the result is given by 
\beq
P(N_{\rm max})\propto N_{\rm max}^{-3}.
\eeq
For a randomly selected inflection point in the landscape, the probability for $N_{\rm max}$ to be in a small range $dN_{\rm max}$ is $P(N_{\rm max}) dN_{\rm max}$.
This conclusion is in agreement with a more heuristic calculation by Yang in Ref.~\cite{Yang}.

\subsection{The attractor region}\label{sec:attractor}
 
Based on the above analysis, we can expect that slow-roll inflation will occur if the shifted field 
\beq
\phi_* \equiv \phi_{1,0} - \sum_a \frac{\rho_{1aa}\phi_{a,0}^2}{2m_a^2} 
\label{phi_*}
\eeq
is in the attractor range (\ref{phi1range}), $-2\kappa H^2 / \abs{\thi} <\phi_* \lesssim H^2/ \abs{\thi}$. 
Thus, even if we start with large values of $\phi_a$ ($\phi_{a,0}\ll\Lambda$), we may still have a range $\Delta\phi_{1,0} \sim 2\kappa H^2/\abs{\rho}$ that gives enough inflation, but now this range is centered at
\beq
\phi_{1,0} \sim \sum_a \frac{\rho_{1aa}\phi_{a,0}^2}{2m_a^2} .
\eeq

\begin{figure}[t] %  figure placement: here, top, bottom, or page
   \centering
   \includegraphics[width=2.5in]{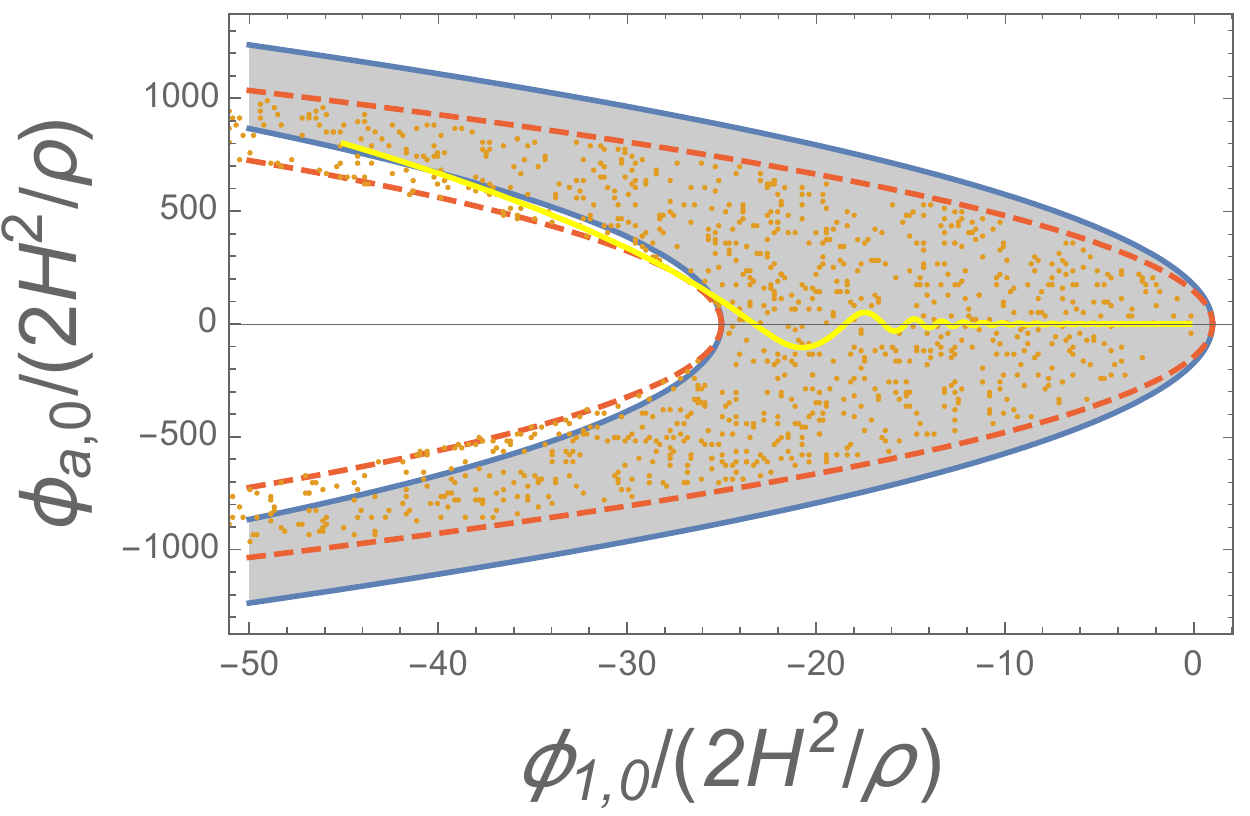} 
   \qquad
   \includegraphics[width=2.5in]{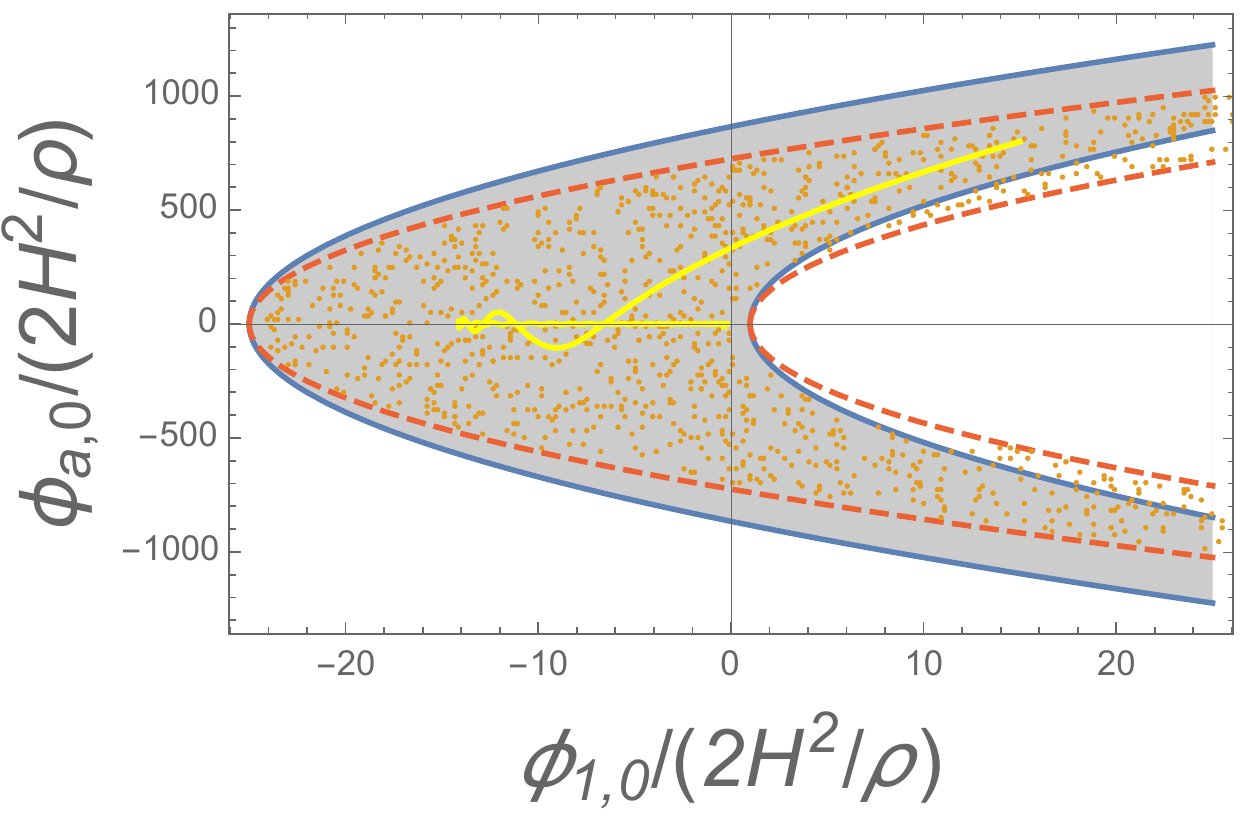} 
   \caption{
   The analytic attractor region of slow-roll inflation for the two-field model indicated in the text
   is shown by grey shading.  The orange dots indicate the attractor region found numerically for the same model.  The agreement between the analytic and numerical results becomes very close after inclusion of an adjustment factor of 0.7, as indicated by dashed red lines.  Left and right panels correspond to $\rho_{1aa}<0$ and $\rho_{1aa}>0$, respectively.  We show examples of two field trajectories as yellow lines.
}      
   \label{fig:attractor}
\end{figure}

As a test of this analysis, we 
numerically solved the field equations for $\phi_1$ and one additional field $\phi_a$: 
\beq
 &&\ddot \phi_1+3 H \dot{\phi}_1 + \grad + \frac{1}{2} \rho \phi_1^2 
 + \thi_{11a} \phi_1 \phi_a 
 + \frac{1}{2} \thi_{1aa} \phi_a^2 = 0. 
\label{eq1}
 \\
 &&\ddot \phi_a+3 H \dot{\phi}_a + m_a^2 \phi_a
 +  \thi_{1aa} \phi_1 \phi_a
 +  \frac{1}{2} \thi_{11a} \phi_1^2  = 0. 
\label{eq2}
\eeq
We assumed $\rho_{11a} = 0$ and $\grad = 0$ for simplicity and took $\rho = - \abs{\rho_{1aa}}  = - U_0 / \Lambda^3$, $m_a = U_0 / \Lambda^2$, and $\Lambda = 0.01$ in this example.  
We show two examples of trajectories in the field space as yellow lines in Fig.~\ref{fig:attractor}, 
where we used $\rho_{1aa} = U_0 / \Lambda^3$ ($-U_0/ \Lambda^3$) and the initial values
$\phi_{1,0}/(2 H^2 / \rho) =  -45$ $(+15)$ in the left (right) panel.  We used  $\phi_{a,0}/(2 H^2 / \rho) = 800$ in both panels.  In both examples $|\phi_{a,0}|$ is much greater than $|\phi_{1,0}|$, but after a few oscillations the oscillation amplitude of $\phi_a$ is strongly damped and the field $\phi_1$ reaches the attractor range (\ref{phi1range}), so that slow-roll inflation can begin.

We chose the initial values $\phi_{1,0}$ and $\phi_{a,0}$ at random 
and marked the choices that led to slow-roll inflation by orange dots in Fig.~\ref{fig:attractor}.%
\footnote{ A range of initial values in a two-field cubic potential model has been studied by Blanco-Pillado {\it et al}  in Ref.~\cite{Jose} to determine the values that lead to slow-roll inflation.  The main difference from our work is that they explored a range of fields $\sim\Delta\phi_{1,0}$ in both $\phi_1$ and $\phi_a$ directions, while we found that the attractor region extends far beyond this range.
}

The region outlined by the orange dots is in a qualitative agreement with the shaded attractor region that we found analytically.  Noting that there is an $\mathcal{O}(1)$ uncertainty in the analytic treatment, we fitted the data by adding an $\mathcal{O}(1)$ factor in the second term on the right-hand side of \eq{phi_*}. We find that inclusion of a factor of $0.7$ leads to a remarkably good agreement with the data, as indicated by the red dashed curves. This shows that the width of the attractor range in the $\phi_1$ direction is indeed given by $\Delta \phi_{1,0} \sim 2 \kappa H^2 / \abs{\rho}$. 

The attractor region can be characterized by the fraction $f$ of volume it occupies in a correlation-length-size region of the landscape, centered at the inflection point.  In a $1D$ landscape, this fraction is 
\beq
f \sim \frac{\Delta\phi_{1,0}}{\Lambda} \sim \kappa\Lambda^2,
\label{f}
\eeq
where $\Delta\phi_{1,0} \sim \kappa\Lambda^3$ is the size of the $1D$ attractor range in Eq.~(\ref{phi1range}).  In our $2D$ model (\ref{eq1})-(\ref{eq2}), the boundaries of the attractor region are two identical parabolas shifted by $\Delta\phi_{1,0}$.  The ratio of the area between these parabolas and the area $\sim \Lambda^2$ of a correlation-length region is still given by Eq.~(\ref{f}).  It is not difficult to see that this also holds in the general multi-dimensional case: the attractor fraction of the correlation-length volume ($\sim \Lambda^N$) in an $N$-dimensional random landscape is given by Eq.~(\ref{f}). 

I-S. Yang assumed in Ref.~\cite{Yang} that the end points of quantum tunneling are more or less uniformly distributed in the field space, in which case the probability of inflation would be proportional to the volume fraction $f$.  Yang offered a heuristic argument that this fraction should decrease exponentially with the number of landscape dimensions $N$.  However, our analysis indicates that, surprisingly, $f$ appears to be independent of $N$.  Furthermore, we shall see in Sec.~\ref{sec:tunneling}  that the assumption of a uniform distribution of tunneling points also needs to be reconsidered.

\section{Comparison with the DBM model}\label{sec:DBM}

We found in the preceding Section that inflation in a small-field random Gaussian landscape is typically one-dimensional.  After a brief period of rapid oscillation, the field settles into a narrow slow-roll track.  This is in contrast with the picture suggested by the Dyson Brownian Motion (DBM) model \cite{Marsh,Dias:2016slx,Freivogel,Marsh2}, asserting that inflation in a large landscape is generically multi-field, with a number of fields having small masses and participating in the slow roll.  We also found no evidence for the rapid steepening of the potential predicted by the DBM model.  The difference between the two approaches can be understood from the following heuristic argument.  

As we mentioned in the Introduction, the DBM process rapidly drives the probability distribution for the Hessian matrix $\zeta_{ij}$ to that of the Gaussian Orthogonal Ensemple (GOE), which is given by \cite{Wigner}
\beq
P(\zeta)\propto \exp(-{\cal Q}) ,
\label{PQ}
\eeq
where
\beq
{\cal Q} = \frac{1}{{\tilde b}^2} {\rm Tr} \zeta^2 
\label{Q-GOE}
\eeq
with certain constant ${\tilde b}$.  On the other hand, the Hessian distribution for a random Gaussian field (RGF) is given, after integration over $U$, by (\ref{PQ}) with \cite{Fyodorov,BrayDean}
\beq
{\cal Q} = \frac{1}{b^2} \left[ {\rm Tr} \zeta^2 -\frac{1}{N+2} ({\rm Tr}\zeta)^2\right] 
\label{Q-RGF}
\eeq
and with $b$ from Eq.~(\ref{b}).

We can represent the eigenvalues of the Hessian as $\lambda_i = {\bar\lambda}+\delta\lambda_i$ ($i=1, ... , N$), where ${\bar\lambda}$ is the average eigenvalue and $\sum_i \delta\lambda_i =0$.  Then
\beq
{\rm Tr}\zeta^2 = \sum_i (\delta\lambda_i)^2 + N{\bar\lambda}^2
\label{GOE}
\eeq
and 
\beq
{\rm Tr}\zeta^2-\frac{1}{N+2} ({\rm Tr}\zeta)^2 = \sum_i (\delta\lambda_i)^2 + \frac{2N}{N+2} {\bar\lambda}^2 \approx  \sum_i (\delta\lambda_i)^2 + 2 {\bar\lambda}^2 ,
\label{RGF}
\eeq
where we assumed $N\gg 1$ in the last step.  
(Note that ${\bar\lambda}=N^{-1}\sum_i \lambda_i$ is the average over a particular realization of the matrix $\zeta$, not the ensemble average.)  

For a generic point in the landscape, the numbers of positive and negative Hessian eigenvalues are about equal, their distribution is approximately symmetric about $\lambda =0$, and ${\bar\lambda}\approx 0$.  
In this case, the GOE and RGF eigenvalue distributions are essentially the same and are given by the Wigner semicircle law (\ref{Wigner}) with ${\bar\lambda}=0$.

The difference between the GOE and RGF ensembles becomes apparent when we compare the coefficients of the ${\bar\lambda}^2$ terms in Eqs.~(\ref{GOE}) and (\ref{RGF}).  In the GOE ensemble, fluctuations of ${\bar\lambda}$ away from zero are very strongly suppressed.  The probability of having ${\bar\lambda}$ comparable to the width of the Wigner distribution -- for example, the probability of having all, or almost all Hessian eigenvalues positive -- is \cite{DeanMajumdar} $P\propto \exp(- {\cal O}(1) N^2)$, while for the RGF it is \cite{Fyodorov,BrayDean} $P\propto \exp(- {\cal O}(1) N)$.  

An inflection-point inflation starts at a rare point in the landscape, where one of the Hessian eigenvalues is very small, while all other eigenvalues are positive.\footnote{The same considerations apply to hilltop inflation, which starts at a point where a one or few eigenvalues are very small and the rest are all positive.}  Then the DBM process rapidly drives the field ${\bm \phi}$ towards regions where some Hessian eigenvalues are negative.  This "evolutionary pressure" is rather strong, because of the strong bias against nonzero ${\bar\lambda}$ in the GOE ensemble.  As positive eigenvalues are "pushed" to the negative side, they cross zero.  Then the corresponding field starts fluctuating, and inflation becomes multi-field.  When some eigenvalues become sufficiently negative, the potential steepens and the slow roll ends. 

As we argued in Section 3, this behavior is not characteristic of inflation in RGF.  Thus, DBM does not seem to provide an adequate description of inflation in a random Gaussian landscape.

\subsection{Comment on a Gaussian correlation function}
 
Here we comment on a special case where the correlation function has a Gaussian form (\ref{correlation function}). The difference from a generic correlation function is apparent 
when we consider the Hessian distribution for a fixed value of $U$. 
For a generic case, it is given by 
\beq
 &&Q = 
 \frac{1}{b^2}  \Tr \lmk \zeta - \lambda_* (U) {\bm 1} \rmk^2
 - \frac{1}{Nb^2 } \lmk 1 - \frac{c}{N} \rmk \lkk \Tr \lmk \zeta - \lambda_* (U) {\bm 1} \rmk \rkk^2 
\label{QGauss}
 \\
 &&\lambda_*(U) = - \frac{\sigma_1^2 }{N \sigma_0^2} (U - \bar{U}), 
\eeq
where $c = {\cal O}(1)$ is determined by the moments. 
However, for a Gaussian correlation function it is given by 
\beq
 Q = 
 \frac{1}{b^2} \Tr \lmk \zeta - \lambda_* (U) {\bm 1} \rmk^2, 
\eeq
because of an accidental cancellation in the coefficient of the second term in (\ref{QGauss}). 
The former one is almost identical to \eq{Q-RGF} 
except for a constant shift, 
while 
the latter one is equivalent to the GOE (\ref{Q-GOE}) 
with a constant shift. 
Therefore, 
in the case of a Gaussian correlation function the distribution of the Hessian is just given by 
the GOE with a constant shift, due to an accidental cancellation.\footnote{ Note, however, that after integration over $U$ the Hessian distribution is given by Eq.~(\ref{Q-RGF}) for any correlation function, including the Gaussian \cite{Fyodorov,BrayDean}.} 
Since this is not a generic case, we do not focus on this case 
but consider a generic correlation function.

\section{Distribution of the initial values}
\label{sec:tunneling}

\subsection{General formalism}\label{sec:initial}

Quantum tunneling that leads to bubble nucleation is described by an $O(4)$-symmetric instanton solution ${\bm \phi} (r)$ of the Euclidean field equations
\bel{instantonEoM}
\frac{d^2\phi_i}{dr^2}+ \frac3r \frac{d\phi_i}{dr}=\frac{dU}{d\phi_i}~
\ee
with suitable boundary conditions.
Here we assume that gravitational effects on the tunneling can be neglected, which is usually the case in a small-field landscape.  The initial values of the fields $\phi_i$ after tunneling are set by their values at the center of the instanton,
\beq
\phi_{i,0} = \phi_i(0).
\eeq

It can be easily verified that Eq.~(\ref{instantonEoM}) does not change its form
under a rescaling
\beq
 &&\phi_i =  \Lambda {\bar\phi_i}
 \label{rescale of phi}
 \\
 &&r =  \Lambda U_0^{-1/2} {\bar r}, 
 \label{rescale of r}
 \\
 &&U({\bm \phi})= U_0 {\bar U}({\bar{\bm \phi}}).
 \label{rescale of U}
\eeq
The rescaled potential ${\bar U}({\bar{\bm \phi}})$ is characterized by the same correlation function as $U({\bm \phi})$, but with $U_0=\Lambda=1$.  In the absence of small parameters, one might expect that the ensemble distribution for the initial values ${\bar\phi}_{i,0}$ would spread over a wide range of size $\sim 1$ in the field space.  The values of $\phi_i$ would then be spread over a range $\sim \Lambda$.
As we already mentioned in Sec.~2.1, this is indeed the case for tunneling to an inflection point in a $1D$ landscape.  However, tunneling to a generic minimum of the potential in $1D$ tends to give a value of $\phi_0$ very close to the minimum \cite{Sarid,Jun}.  These features of $1D$ tunneling suggest that the field distribution may be much narrower in the directions orthogonal to that of the zero eigenvalue.  In the next subsection we will show that this is indeed the case.

\subsection{Initial values for multi-filed tunneling}

The instanton solution ${\bm \phi}(r)$ of Eq.~(\ref{instantonEoM}) describes the motion of the field ${\bm \phi}$ in the upside-down potential $-U({\bm \phi})$, with $r$ playing the role of time.  The field starts at 
zero velocity with ${\bm \phi}(r=0)={\bm \phi}_0$ and approaches the false vacuum value at $r\to\infty$.  In a generic configuration, the false vacuum is displaced from the inflection point (${\bm \phi}=0$) by $\sim \Lambda$, both in $\phi_1$ and $\phi_a$ directions.

We shall consider an instanton whose center is relatively close to the inflection point, $\abs{{\bm \phi}_0}\ll \Lambda$.  In the vicinity of the inflection point, Eq.~(\ref{instantonEoM}) can be approximated as
\beq
 \frac{\dd^2 \phi_1}{\dd r^2} + \frac{3}{r} \frac{\dd \phi_1}{\dd r} = 
 \eta + \frac{\rho}{2} \phi_1^2, 
 \label{separateEoM1}
 \\
 \frac{\dd^2 \phi_a}{\dd r^2} + \frac{3}{r} \frac{\dd \phi_a}{\dd r} = 
 m_a^2 \phi_a + \frac{\rho_{11a}}{2} \phi_1^2.
 \label{separateEoM2}
\eeq
Here we assumed that $\abs{\phi_a} \ll \abs{\phi_1}$, which will be justified {\it a posteriori}. 
We shall also neglect the term $\eta$ in the equation for $\phi_1$; this term will be taken into account later.  

\begin{figure}[t] %  figure placement: here, top, bottom, or page
   \centering
   \includegraphics[width=2.5in]{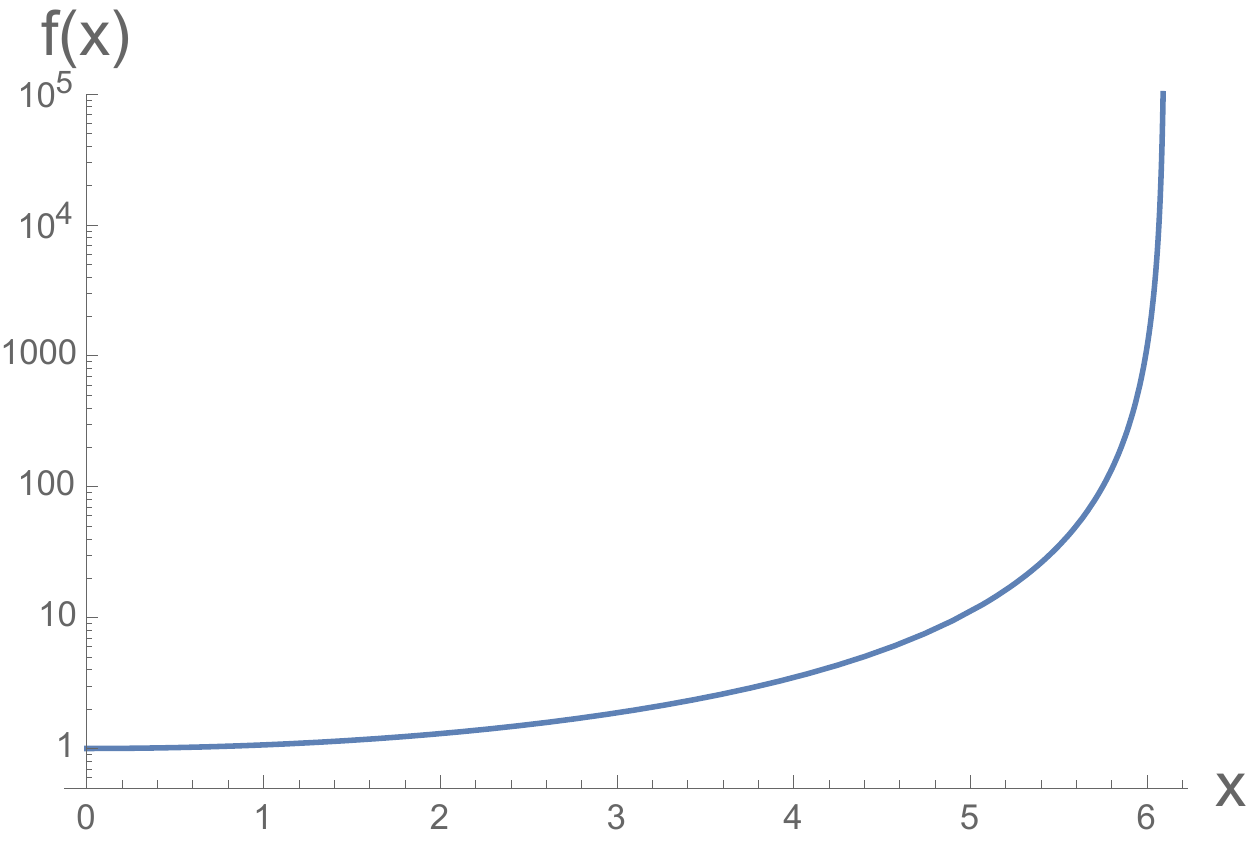} 
   \caption{
 Plot of $f(x)$. 
   }
   \label{fig:fx}
\end{figure}

With these approximations, $\phi_1(r)$ can be represented as 
\beq
 \phi_1 (r) =  \phi_{1,0} f \lmk \sqrt{\thi \phi_{1,0}} \, r \rmk, 
 \label{phi_1}
\eeq
where $f (x)$ is a function satisfying 
\beq
 f'' + \frac{3}{x} f' = \frac{f^2}{2} 
\eeq
with boundary conditions $f(0)=1$, $f'(0)=0$.
The solution is shown in Fig.~\ref{fig:fx}, where we can see that $f(x)$ diverges at $x \simeq 6.1$.  This solution becomes inaccurate when $\phi_1$ reaches values $\sim \Lambda$.
For small values of $x$,
\beq
f(x) = 1 + x^2 /16 + \mathcal{O}(x^3).
\label{smallx}
\eeq

The initial bubble radius $r_0$ can be estimated as the value of $r$ at which $\phi_1$ significantly deviates from its value $\phi_{1,0}$ at the bubble center:
\beq
 r_0 \simeq \frac{c_1}{\sqrt{\thi \phi_{1,0}}}, 
 \label{phi0 third}
\eeq
where $c_1 \sim 5$.  Let us compare this with the generic bubble wall thickness $\delta \sim \Lambda/ \sqrt{U}$.  With $\thi\sim U/\Lambda^3$, we have $r_0/\delta \sim c_1(\Lambda/\abs{\phi_{1,0}})^{1/2} \gg 1$ for $\abs{\phi_{1,0}}\ll \Lambda$.  This means that tunneling close to the inflection point requires a thin-wall bubble of radius much larger than the wall thickness.  This typically requires that the potential at the inflection point should be nearly degenerate with that of the false vacuum.\footnote{For a thin-wall bubble, the instanton solution stays near $\phi_{1,0}$ for a long Euclidean time, so that the friction term in Eq.~(\ref{instantonEoM}) becomes unimportant.  Then energy is approximately conserved, and the potential at the endpoints of instanton solution is nearly the same.}  We note also that for $|\phi_{1,0}|\sim \kappa\Lambda^3$ we have $r_0\sim U^{-1/2} \sim H^{-1}$, in which case gravitational effects on tunneling may be important.  We will not attempt to analyze these effects here.

Next, we consider Eq.~(\ref{separateEoM2}) for $\phi_a$.  
Let us first consider 
a particular solution $\phi_a^{(p)}$ 
that includes the effect of the interaction term $\rho_{11a} \phi_1^2$, 
where $\phi_1$ is given by \eq{phi_1}. 
It is 
\beq 
 \phi_a^{(p)} (r) \simeq 
 -  \frac{\rho_{11a} \phi_{1,0}^2}{2 m_a^2} 
 + 
 \frac{\rho_{11a} \phi_{1,0}^2}{16m_a^2} \rho \phi_{1,0} r^2
 + \dots, 
 \label{phi-p}
\eeq
for $\phi_{1,0} \ll \Lambda$, where the dots represent higher order terms in terms in $\sqrt{\rho \phi_{1,0}} \, r$. 
We neglect the second and higher-order terms in what follows because we are interested in the case where 
$\sqrt{\rho \phi_{1,0}} \, r \lesssim 1$. 
The solution of the homogeneous equation 
$\phi_a^{(h)}$ is given by 
\beq
 \phi_a^{(h)} (r) = c_0 \frac{2 J_1 (i m_a r)}{i m_a r}, 
\eeq
where $c_0$ is a constant. 
The solution of \eq{separateEoM2} can be thus written as 
\beq
 \phi_a (r) 
 &=&  \phi_a^{(p)} (r) +  \phi_a^{(h)} (r) 
 \\
 &\simeq& -  \frac{\rho_{11a} \phi_{1,0}^2}{2 m_a^2} + 
 \lmk \phi_{a,0} -  \frac{\rho_{11a} \phi_{1,0}^2}{2 m_a^2} \rmk \frac{2 J_1 (i m_a r)}{i m_a r}, 
\eeq
where $\phi_{a,0}$ ($\equiv \phi_a (0)$) is the tunneling endpoint of $\phi_a$. 
The asymptotic form of the Bessel function is given by 
\beq
 \frac{J_1 (im_a r) }{i m_a r} \sim \frac{e^{m_a r}}{\sqrt{2 \pi (m_a r)^3}}, 
\eeq
for $m_a r \gg 1$.  We see that $\phi_a(r)$ grows exponentially, but consistency requires that it should remain sufficiently small ($\ll\Lambda$) until $r \sim r_0$.  It follows that the initial value $\phi_{a,0}$ should satisfy
 \beq
 && \abs{ \phi_{a, 0} -  \frac{\rho_{11a} \phi_{1,0}^2}{2 m_a^2} } \lesssim \Lambda (m_a r_0)^{3/2} e^{-m_a r_0}, 
 \label{phi0 second}
\eeq
where we assume $m_a r_0 \gg 1$.  Using \eq{phi0 third}, this can be rewritten as
\beq
 \abs{ {\phi_{a,0}} -  \frac{\rho_{11a} \phi_{1,0}^2}{2 m_a^2} } \lesssim \Lambda \lmk \frac{c_1^2 m_a^2}{\rho \phi_{1,0}} \rmk^{3/4} 
 \exp \lkk - \frac{c_1 m_a}{\sqrt{\rho \phi_{1,0}}} \rkk. 
 \label{phi_0 relation}
\eeq
Thus the tunneling endpoint of $\phi_a$ is exponentially close to 
$\rho_{11a} \phi_{1,0}^2 / 2 m_a^2$, which is much smaller than 
$\phi_{1,0}$ for $|\phi_{1,0}|\ll\Lambda$.

Finally, we comment on the effect of the $\eta$ term in \eq{separateEoM1}.  At small values of $r$, this equation can be approximated as
\beq
 \frac{\dd^2 \phi_1}{\dd r^2} + \frac{3}{r} \frac{\dd \phi_1}{\dd r} = 
 \eta + \frac{\rho}{2} \phi_{1,0}^2. 
\eeq
The solution is
\beq
 \phi_1 (r) = \phi_{1,0} + \frac{1}{16} \rho\phi_{1,0}^2 r^2 + \frac{1}{8} \eta r^2 . 
\eeq
The last term is negligible compared with the second one if
\beq
\abs{\phi_{1,0}}\gg (\eta/\rho)^{1/2}\sim \frac{U}{\rho N_{\rm max}} \sim \frac{\Lambda^3}{N_{\rm max}},
\label{upper bound}
\eeq
where we have used Eq.~(\ref{Nmax}) for $N_{\rm max}$.  This condition is satisfied unless $\phi_{1,0}$ is extremely close to the inflection point.

\subsection{Numerical results for a toy model}\label{sec:toy model}

Here we consider a numerical example to check the analysis in Sec.~5.2, in particular the relation \eq{phi_0 relation}.  We consider a two-dimensional mini-landscape with the potential of the form
\beq
 U(\phi_1, \phi_a) &&= \lmk \eta \phi_1 + \frac{m^2}{2} \phi_1^2 + \frac{\rho}{6} \phi_1^3 
 + \frac{m_a^2}{2} \phi_a^2 
 + \frac{\rho_{aaa}}{6} \phi_a^3 
 + \frac{\rho_{1aa}}{2} \phi_1 \phi_a^2 
 + \frac{\rho_{11a}}{2} \phi_1^2 \phi_a 
 \rmk 
 \nonumber\\
 &&\quad \times \lkk \frac{m_1^2}{2} \lmk \phi_1 - R \cos \theta \rmk^2 + \frac{m_2^2}{2} \lmk \phi_a - R \sin \theta \rmk^2 
 + \delta \rkk, 
 \label{toy model}
\eeq
where $m$, $m_1$, $m_2$, $R$, $\theta$, $\delta$ are constant parameters. 
For $m=0$, there is an inflection point at $\phi_1 = \phi_a = 0$.
The parameters $R$ and $\theta$ determine 
the location of false vacuum, 
and $\delta$ determines its height. 

\begin{figure}[t] %  figure placement: here, top, bottom, or page
   \centering
   \includegraphics[width=5in]{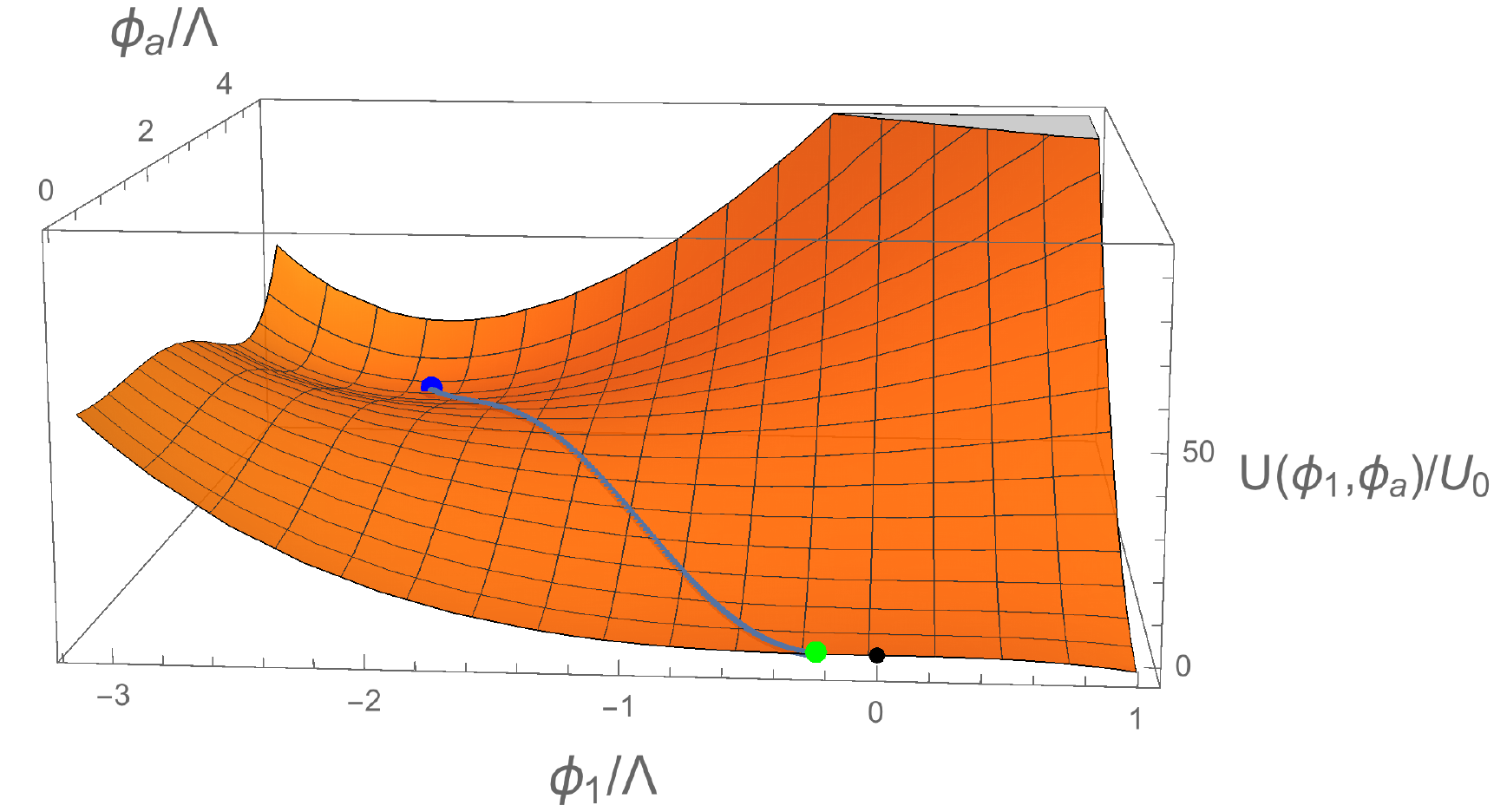} 
   \caption{
An example of the potential (\ref{toy model}), where we use the parameters indicated in the text with $\theta = 2\pi / 3$ and $\delta /U_0= 0.5$.  This choice of parameters was mostly made for the clarity of presentation in the figure. 
The false vacuum and inflection point are marked by blue and black dots, respectively. 
The gray line shows the instanton trajectory, and the green dot is the tunneling point. 
   }
   \label{fig:potential}
\end{figure}

We take $m_1 = m_2 = m_a = U_0 / \Lambda^2$, 
$\rho = \rho_{aaa} = 3 \rho_{1aa} = 3 \rho_{11a} = -U_0/ \Lambda^3$, $m = \eta = 0$, and $R = 5 \Lambda$ as an example, 
while 
we choose $\theta$ and $\delta / U_0$ randomly within the ranges of $(\pi /2,\pi)$ and $(0,1)$, respectively.   An example of the potential is shown in Fig.~\ref{fig:potential}, 
where the false vacuum and the inflection point are marked by blue and black dots, respectively. 

We discard the realizations where 
there is no false vacuum near $(\phi_1, \phi_a) = (R \cos \theta,  R \sin \theta)$, 
which is sometimes the case for $\delta  / U_0 \gtrsim 0.8$.  Note that the parameters of the landscape $\Lambda$ and $U_0$ can be eliminated by rescaling of the variables [see \eq{rescale of phi}], so the results below are independent of $\Lambda$ and $U_0$.

\begin{figure}[t] %  figure placement: here, top, bottom, or page
   \centering
   \includegraphics[width=5in]{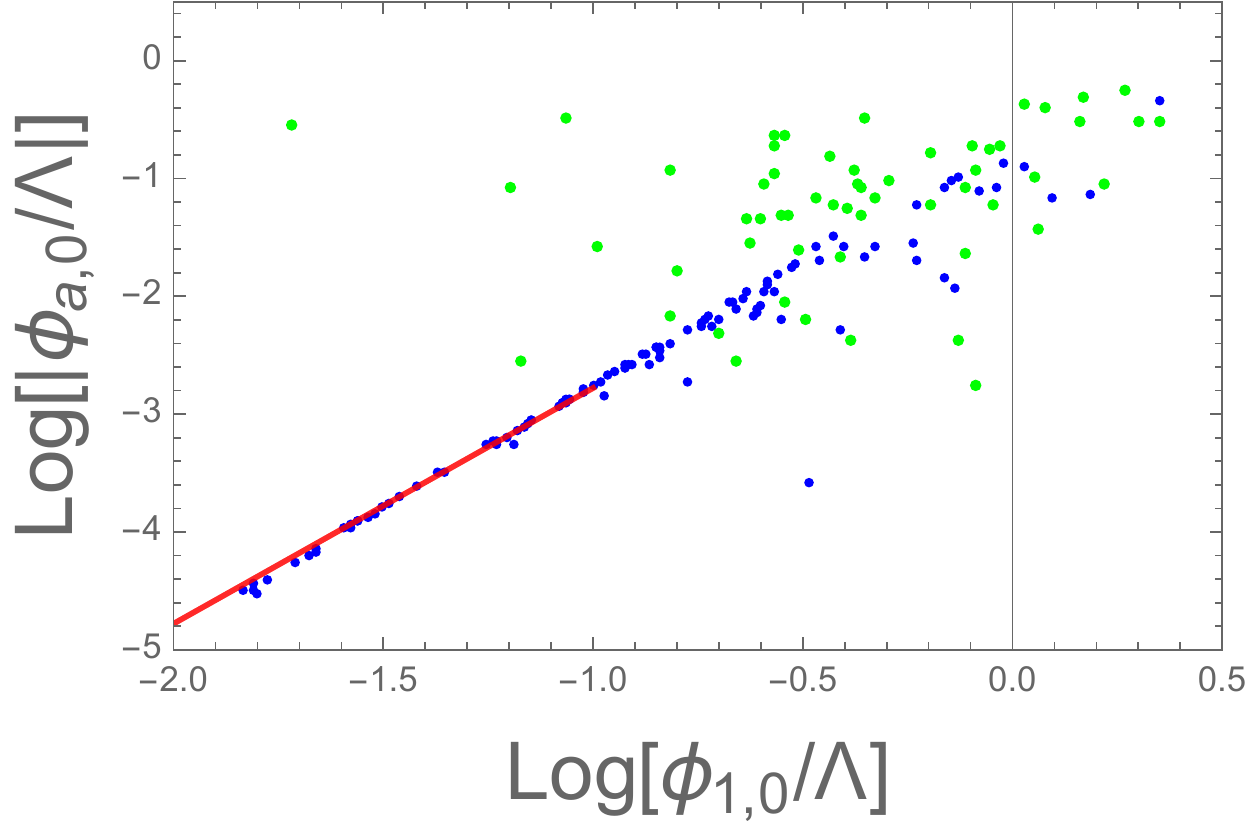} 
   \caption{
 Distribution of tunneling points in the toy model (\ref{toy model}). 
 We choose $\theta$ and $\delta / U_0$ randomly within the ranges of $(\pi/2,\pi)$ and $(0,1)$, respectively. 
 The blue dots represent the case where $\phi_{a,0} < 0$ 
 while the green ones represent the case where $\phi_{a,0} > 0$. 
 The red line represents $\rho_{11a} \phi_{1,0}^2 / 2 m_a^2$ with $\phi_{1,0} / \Lambda < 0.1$, 
 which is the analytic formula (\ref{phi_0 relation}) that is valid for $\phi_{1,0} / \Lambda \ll 1$. 
   }
   \label{fig:initial}
\end{figure}

We found the instanton solution using the efficient algorithm of Ref.~\cite{Ali} and determined 
the tunneling point for each realization. 
The resulting distribution is shown in Fig.~\ref{fig:initial}, 
where 
the blue dots represent the case where $\phi_{a,0} < 0$ 
while the green ones represent the case where $\phi_{a,0} > 0$. 
We see that green dots are rarer for smaller $\phi_{1,0}$ 
and the blue dots tend to be close to $\rho_{11a} \phi_{1,0}^2 / 2 m_a^2$, 
which is plotted as the red line for $\phi_{1,0} / \Lambda < 0.1$ ($\ll 1$). 
The plot shows that the tunneling points typically concentrate along the flat direction ($\phi_1$-axis), with
$\phi_{a,0}$ close to $\rho_{11a} \phi_{1,0}^2 / 2 m_a^2$ 
when $\phi_{1,0}$ is much smaller than $\Lambda$. 
This result is in a good agreement with the analytic formula (\ref{phi_0 relation}).

Combining this result with that for the inflationary attractor region, 
we find that 
inflation is possible only if the tunneling point is very close to the $\phi_1$-axis, with $\phi_{1,0}$ in the range (\ref{phi1range}) except for some rare realizations.  In other words, the initial conditions have to be very close to those for $1D$ inflation, with $\phi_{a,0}\approx 0$ and with $\phi_{1,0}$ in the $1D$ attractor range.  Although there are some exceptions, where the false vacuum is close to the $\phi_1$ axes and $\delta$ is relatively small, those realizations are rarer for smaller $\phi_{1,0}$.   Then the dynamics remains essentially one-dimensional all the way from bubble nucleation till the end of the slow roll.

\section{Saddle point inflation}\label{sec:saddle}

Inflation in the vicinity of a saddle point of the landscape can be analyzed in much the same way as inflection-point inflation, with similar conclusions.  In this case we have $\eta_i =0$ in the cubic expansion 
(\ref{potential}) of the potential.  For a slow-roll inflation, one of the eigenvalues of the Hessian has to be small and negative, and we can choose the basis in the $\phi$-space so that this eigenvalue corresponds to $\phi_1$ direction.  All other eigenvalues, as well as the coefficients of the cubic expansion $\rho_{ijk}$ will typically have their generic values.  We shall denote the small eigenvalue $-m^2$ and require $m\lesssim H = (U/3)^{1/2}$.\footnote{The Hessian may have several small eigenvalues, but such saddle points will be rare in the landscape.}

The potential near the saddle point has the form of a flat hilltop surrounded by steep rising slopes.  As before, inflation is approximately one-dimensional.  The shape of the potential in the $\phi_1$-direction is
\beq
U(\phi_1)=U-\frac{1}{2}m^2 \phi_1^2 +\frac{1}{6}\rho\phi_1^3 ,
\eeq
where $\rho\equiv \rho_{111}$ and we assume $\rho<0$.  Note that this potential has a shallow local minimum at $\phi_1 =2m^2 /\rho$.

A characteristic feature of hilltop inflation is that it is eternal in the range $\abs{\phi_1}\lesssim U^{3/2}/m^2 \equiv\phi_q$, where the field $\phi_1$ undergoes quantum diffusion \cite{AV83}.  The slow-roll regime corresponds to $\phi_q\lesssim \phi_1\lesssim \phi_{\rm end}$, where $\phi_{\rm end}$ is determined by the condition $\abs{U''}/U\sim 1$, $\phi_{\rm end}\sim U/\abs{\rho} \sim \Lambda^3$.  The number of e-folds during the slow roll is bounded by
\beq
N_{\rm max} = \int_{\phi_q}^{\phi_{\rm end}} d\phi_1 \frac{U(\phi_1)}{U'(\phi_1)} \approx \frac{U}{m^2} \ln \left(1+\frac{2m^2}{\rho\phi_q}\right) .
\eeq
The logarithm here is $\lesssim 100$; hence we need $m\lesssim H$.

If the initial conditions after tunneling are such that $\phi_a\approx 0$, the attractor region in this case consists of two segments, separated by a large gap [$\phi_{1,0} \sim (- \kappa \Lambda^3, 0)$] where the field ends up on the "wrong" side of the hill and rolls into the shallow minimum. The attractor range of inflation is thus in the intervals $\Delta\phi_{1,0} \sim \Lambda^3$ near the boundaries of this range (i.e., $\phi_{1,0} \sim - \kappa \Lambda^3$ and $\sim 0$). If, on the other hand, $\abs{\phi_a}\gg \kappa \Lambda^3$, then essentially the same analysis as in Sec.~3.2 leads to the conclusion that $\phi_1$ is shifted by the amount (\ref{shift}) shortly after the bubble nucleation.  The resulting attractor region consists of two parts having the same horseshoe shape as in the inflection-point case.  The difference is that the widths of the horseshoes and their volume are smaller by a factor of $1/\kappa \simeq 0.04$).

The distribution of tunneling points is also expected to be similar.
Since the mass term in the $\phi_1$ direction is much smaller than its typical value, 
$\abs{m^2}\ll U_0/\Lambda^2$,
we expect that the instanton solution is not sensitive to its magnitude, so the resulting distribution is close to the one we found in Sec.~\ref{sec:toy model} for $m=0$.  We verified this numerically for the toy model
(\ref{toy model}) with $m^2 = 0.01~ U_0/\Lambda^2$ and all other parameters the same as in Sec.~\ref{sec:toy model}.  As before, we found that the tunneling points concentrate along the flat direction
Thus we conclude again that the inflationary dynamics is effectively one-dimensional after tunneling.

\section{Conclusions and discussion}\label{sec:conclusion}

In this paper we studied slow-roll inflation in large random Gaussian landscapes.  We assumed the landscape to be small-field, with the correlation length $\Lambda$ much smaller than the Planck scale, $\Lambda\ll 0.1$.  In this case inflation typically occurs in small patches of the landscape, localized near saddle or inflection points, so the potential can be accurately approximated by Taylor expansion about these points up to cubic order.  Our main conclusions can be summarized as follows.  

{\it (i)} Inflation in this kind of landscape is approximately one-dimensional, with the field moving in a nearly straight line during the slow roll.  

{\it (ii)} We defined the attractor range of inflation as the set of initial values of the scalar fields ${\bm \phi}$ that lead to slow roll.  This range can be characterized by the fraction of volume $f_N$ it occupies in the $N$-dimensional correlation-length-sized region centered at the corresponding saddle or inflection point.  In a $1D$ landscape, $f_1$ is comparable to the fraction of the correlation length where the slow-roll conditions are satisfied, $f_1\sim \Lambda^2$ \cite{MVY2}.  Naively, one might expect that in a large landscape $f_N$ decreases exponentially with $N$ \cite{Yang}.  We found, however, that, surprisingly, $f_N$ is nearly independent of $N$, $f_N\sim f_1$.  When the field ${\bm \phi}$ starts relatively far from the slow-roll region, it undergoes rapid damped oscillations in the directions orthogonal to the inflationary track, and cubic interaction terms cause a large shift of the field along the track, so that it may end up in the slow-roll region.  The resulting attractor range stretches far beyond the slow-roll regime, as illustrated in Fig.~\ref{fig:attractor}.

{\it (iii)} The probability of inflation would be proportional to the attractor volume fraction if the tunneling endpoints were uniformly distributed through the landscape.  However, we found this not to be the case.  Our study of the instantons, both analytical and numerical, indicates that the tunneling endpoints tend to concentrate along the flat direction.  If the endpoints spread more or less uniformly along this line,  
the probability of inflation would still be proportional to $f_1$.  A quantitative analysis of this issue would require a statistical study of tunneling in the landscape, which we have not attempted here.

{\it (iv)} Our picture of inflation in a large landscape is rather different from that suggested by the Dyson Brownian Motion (DBM) model in Refs.~\cite{Marsh,Dias:2016slx,Freivogel,Marsh2}.  In particular, we find no evidence for the rapid steepening of the potential and for the resulting suppression of the number of inflationary e-folds predicted in this model. On the contrary, we find that the distribution for the number of e-folds is the same as in the $1D$ case, in agreement with Ref.~\cite{Yang}.  

The DBM model uses an expansion of the potential up to quadratic terms and assumes that the evolution of the Hessian matrix $\zeta_{ij}=\partial^2 U/\partial \phi_i \partial\phi_j$ along the inflationary path is described by the Dyson stochastic process.  We see, however, no reason to expect this description to be accurate in a random Gaussian landscape.  Inflation occurs in a small patch of the landscape, so we can use Taylor expansion.  The first and some of the second derivatives of the potential in this patch are small, but the third derivatives are not; hence expansion up to cubic terms should be adequate.  The resulting cubic potential does not vary stochastically along a smooth path and does not exhibit any steepening (other than a cubic steepening, as in the $1D$ case).  On the other hand, the DBM process is known to drive the Hessian spectrum towards negative values.  This may explain the steepening, as well as the appearance of low-mass modes (with inflation becoming multi-field).

An important limitation of our analysis is that we studied the probability of inflation in the sense of "probability in the landscape". In other words, for a randomly selected inflection or saddle point in the landscape, we discussed the probability that the potential in the vicinity of that point can support slow-roll inflation.  This is rather different from the probability that this kind of inflation has actually happened in our past.  Calculation of the latter quantity would require accounting for various anthropic factors, as well as some choice of measure on the multiverse.  We expect to return to this issue in a separate publication.

A random Gaussian landscape is, of course, just a simple model.  It may give some useful insights, but eventually one hopes to investigate more realistic landscape models, as it was done, for example, in Refs. \cite{Baumann, Jose, Linde:2016uec}.

\section{Acknowledgement}
We are grateful to Jose Blanco-Pillado for useful discussions.
This work is supported by the National Science Foundation under grant 1518742. M.Y. is
supported by the JSPS Research Fellowships for Young Scientists.

\appendix

\section{Integrals of Bessel functions}\label{Bessel}

Here we calculate the integral involving a product of Bessel functions that we used 
to derive \eq{Bessel result}. 

We first quote a formula (5.55) from Ref.~\cite{GR}: 
\beq
\int\frac{dx}{x} J_p(\alpha x)J_q(\alpha x) = \alpha x\frac{J_{p-1}(\alpha x)J_{q}(\alpha x) -J_{p}(\alpha x) J_{q-1}(\alpha x)}{p^2 -q^2} -\frac{J_p(\alpha x)J_q(\alpha x)}{p+q}. 
\label{integral}
\eeq
The integral of the second term in (\ref{y2}) is found by setting $p=0,~ q=2$ in this formula. 
Using the identities $J_{-1}(z)=-J_1(z)$ and
\beq
J_0(z)+J_2(z)=\frac{2}{z} J_1(z),
\eeq
we can write the result as
\beq
\int_0^z \frac{dz'}{z'}J_0(z')J_2(z') =\frac{1}{2} \left[J_1^2(z)-J_0(z)J_2(z)\right].
\label{int2}
\eeq

The integral of the first term in (\ref{y2}) corresponds to $p=q=1$, which makes the right-hand side of Eq.~(\ref{integral}) ill-defined.  To get around this problem, we use the relations
\beq
 J_1 = - \frac{\dd}{ \dd x} J_0 ,  
 \\
 \frac{\dd}{\dd x} \lmk \frac{J_1 (x)}{x} \rmk = - \frac{J_2 (x)}{x} ,
\eeq
and after integrating by parts obtain
\beq
 \int_0^z \frac{dz'}{z'} J_1^2(z') = - \int_0^z \frac{\dd z'}{z'} J_0 (z') J_2(z') 
 - \frac{J_0(z) J_1(z)}{z} + \frac{1}{2} = \frac{1}{2}\left( 1-J_0^2(z) -J_1^2(z)\right) , 
\label{int1}
\eeq
where in the last step we used \eq{int2} and the relation 
\beq
\frac{J_1(z)}{z}=\frac{1}{2}\left(J_0(z)+J_2(z)\right).
\eeq

Now, combining Eq.(\ref{y2}) with (\ref{int2}), (\ref{int1}) we find
\beq
\phi_1(t)=-\frac{\rho_{1aa}\phi_{a,0}^2}{2m_a^2} \left[1-J_0^2(m_a t) -2J_1^2(m_a t) +J_0(m_a t)J_2(m_a t) \right] ,
\label{Bessel app}
\eeq
which is \eq{Bessel result}.

\section{Distribution for the number of e-folds}\label{sec:N_e}

Up to a normalization factor, the distribution for the maximal number of e-folds at an inflection point can be expressed as
\beq
P(N_{\rm max})&&\propto\int dU  \prod_i d\eta_i \prod_i d\lambda_i \prod_{ijk} d\rho_{ijk} 
\nonumber\\
&&\quad 
\times J(\lambda) e^{-{\cal Q}} 
\delta(\lambda_1)|\rho| \left( \prod_{a=2}^N \delta(\eta_a) |\lambda_a| \right) \delta\left(N_{\rm max} -\frac{\pi\sqrt{2} U}{\sqrt{\eta\rho}}\right).
\label{PN}
\eeq
Here, $\lambda_i$ are the eigenvalues of the Hessian, $J(\lambda)$ is the Jacobian transforming from integration over Hessian components $\zeta_{ij}$ to integration over its eigenvalues, and we have used Eq.~(\ref{Nmax}) for $N_{\rm max}$.  As before, we use the notation $\lambda_1$ for the eigenvalue that vanishes at the inflection point and $\rho\equiv\rho_{111}$.  The delta functions with compensating factors  
$|\rho|$ and $|\lambda_a|$ select inflection points where $\lambda_1=0$ and $\eta_a = 0$ for $a=2, ... , N$.  

The exponent ${\cal Q}$ in (\ref{PN})  is
\beq
{\cal Q}={\cal Q}_1(U,\lambda)+{\cal Q}_2(\eta,\rho),
\eeq
where ${\cal Q}_1(U,\lambda)$ depends only on $U$ and $\lambda_i$ and ${\cal Q}_2(\eta,\rho)$ is given by \cite{MVY1}\footnote{The notation we use here is different from that in Ref.~\cite{MVY1}.}
\beq
{\cal Q}_2(\eta,\rho) = A\eta_i \eta_i +B\eta_i \rho_{ijj}  +C \rho_{iik}\rho_{jjk} +D \rho_{ijk}\rho_{ijk},
\eeq
with summation over repeated indices. The coefficients $A,B,C,D$ can be expressed in terms of the moments of the correlation function; they can be estimated as $A\sim\Lambda^2/U_0^2$, $B\sim \Lambda^4/NU_0^2$, $C\sim \Lambda^6/NU_0^2$, $D\sim \Lambda^6/U_0^2$.  The specific forms of ${\cal Q}_1(U,\lambda)$ and of the Jacobian $J(\lambda$) can be found, e.g., in Refs.~\cite{Fyodorov,BrayDean}; we shall not need them here.

After integration over $\eta_i$, we have
\beq
P(N_{\rm max})\propto N_{\rm max}^{-3} \int dU U^2 \prod_i d\lambda_i \prod_{ijk} d\rho_{ijk} J(\lambda) e^{-{\cal Q}} \delta(\lambda_1) \prod_{a=2}^N |\lambda_a| ,
\eeq
with ${\cal Q}_2$ in the exponent now replaced by
\beq
{\cal Q}_2=\frac{A}{\rho^2}\left(\frac{2\pi^2 U^2}{N_{\rm max}^2}\right)^2 +B \frac{\rho_{1jj}}{\rho} \frac{2\pi^2 U^2}{N_{\rm max}^2} +C \rho_{iik}\rho_{jjk} +D \rho_{ijk}\rho_{ijk}
\label{Q2}
\eeq
The factors $N_{\rm max}^{-3}$ and $U^2$ come from integrating the last delta function in (\ref{PN}) over $\eta$.  The effect of the first and second terms in Eq.~(\ref{Q2}) is to suppress integration over very small values of $\rho\lesssim (\lambda^4/N_{\rm max}^2)(U_0/\Lambda^3)$, which contribute very little to the integral.  (The main contribution comes from $|\rho|\sim U_0/\Lambda^3$.)  After dropping these terms the integral becomes independent of $N_{\rm max}$, and thus we obtain
\beq
P(N_{\rm max})\propto N_{\rm max}^{-3}.
\eeq

\end{document}